\documentclass{ws-ijmpa-blank-new}

\usepackage[super,compress]{cite}
\usepackage{graphicx}

\usepackage{amssymb,epsf,latexsym,amssymb,amsmath,mathrsfs}
\usepackage[english]{babel}

\newcommand{\beq}{\begin{equation}}
\newcommand{\eeq}{\end{equation}}
\newcommand{\beqa}{\begin{eqnarray}}
\newcommand{\eeqa}{\end{eqnarray}}
\newcommand{\bsubeqs}{\begin{subequations}}
\newcommand{\esubeqs}{\end{subequations}}

\makeatletter
\@addtoreset{equation}{section}
\renewcommand{\theequation}{\thesection.\arabic{equation}}
\makeatother

\begin{document}
\markboth{F.R. Klinkhamer}
{Lorentz-violating neutral-pion decays}

%
\catchline{}{}{}{}{}
%

\title{\vspace*{-11mm}
Lorentz-violating neutral-pion decays in isotropic\\ 
modified Maxwell theory}

\author{F.R. Klinkhamer}

\address{Institute for Theoretical Physics, Karlsruhe Institute of
Technology (KIT),\\ 76128 Karlsruhe, Germany\\
frans.klinkhamer@kit.edu}

\maketitle


\begin{abstract}
We consider an extension of the Standard Model
with isotropic nonbirefringent Lorentz violation in the photon sector
and restrict the discussion to
the case of a ``fast'' photon with a phase velocity
larger than the maximum attainable velocity of the quarks and leptons.
With our conventions, this case corresponds to a negative
Lorentz-violating parameter $\kappa$ in the action.
The decay rate of a neutral pion into two nonstandard
photons is calculated
as a function of the 3-momentum of the initial pion and
the negative Lorentz-violating parameter $\kappa$ of the final photons.
\end{abstract}
\vspace*{.0\baselineskip}
{\footnotesize
\vspace*{.25\baselineskip}
\noindent \hspace*{5mm}
\emph{Journal}: Mod. Phys. Lett. A \textbf{33} (2018) 1850104
\vspace*{.25\baselineskip}
\newline
\hspace*{5mm}
\emph{Preprint}: arXiv:1610.03315 
}
\vspace*{-5mm}\newline
\keywords{Lorentz invariance, quantum electrodynamics,	
cosmic-ray interactions}
\ccode{PACS Nos.: 11.30.Cp, 12.20.-m, 13.85.Tp, 98.70.Sa}


\section{Introduction}
\label{sec:Introduction}

In this short paper, we calculate the decay rate of a particular
Lorentz-violating process, neutral-pion decay into two photons,
which can be tested by modeling and observing
extensive air showers in the upper Earth's atmosphere.

\section{Theory}
\label{sec:Theory}

We restrict our attention to the strong and electromagnetic
interactions of the Standard Model. The corresponding theory
is a vector-like (nonchiral) $SU(3)\times U(1)$ gauge theory.
The action density of this gauge theory is now
augmented by a single Lorentz-violating CPT-even dimension-4 photonic
term~\cite{ChadhaNielsen1983,KM2002,KS2008,KS2010}
with a dimensionless real coupling constant $\kappa$.
The Lorentz-violating parameter $\kappa$ is
taken to be negative and to have a very small magnitude
(a possible underlying theory has been suggested in
Ref.~\citen{KlinkhamerSchreck2011}).

A photon of 3-momentum $\vec{k}$ then has a modified dispersion relation,
\beq\label{eq:disp-rel-effective-mass-square} 
\left[\omega(\vec{k}\,)\right]^2
= \frac{1-\kappa}{1+\kappa}\;c^2\,|\vec{k}\,|^2\,,
\eeq
and modified polarization 3-vectors,
\bsubeqs\label{eq:polarization-vectors}
\beqa
\vec{e}^{\,(1)}(\vec{k}\,) &=&
\sqrt{\frac{1}{1+\kappa}}\; \mathcal{R}_{\widehat{k}} \cdot
\left(
  \begin{array}{c}
    0 \\
    1 \\
    0 \\
  \end{array}
\right)\,,\\[2mm]
\vec{e}^{\,(2)}(\vec{k}\,) &=&
\sqrt{\frac{1}{1+\kappa}}\; \mathcal{R}_{\widehat{k}} \cdot
\left(
  \begin{array}{c}
    0 \\
    0 \\
    1 \\
  \end{array}
\right)\,,
\eeqa
\esubeqs
with the $3 \times 3$ rotation matrix $\mathcal{R}_{\widehat{k}}$
which transforms the unit column 3-vector $(1,0,0)^\text{T}$ to
the unit column 3-vector
$\widehat{k} \equiv \vec{k}/k \equiv \vec{k}/|\vec{k}\,|$.
See Ref.~\citen{KS2010} for further details.
In addition, we employ the Minkowski metric
$g_{\mu\nu}(x)
=\eta_{\mu\nu}\equiv [\text{diag}(+1,\,-1,\,-1,\,-1)]_{\mu\nu}$
and, from now on, use natural units with $\hbar=c=1$.

Incidentally, the velocity $c$ appearing on the right-hand side
of \eqref{eq:disp-rel-effective-mass-square} corresponds to the
maximum attainable velocity of the quarks and leptons.
See also the last paragraph of Sec.~\ref{sec:Discussion}
for a brief discussion of the theory considered.

\section{Calculation}
\label{sec:Calculation}

In the theory as outlined in Sec.~\ref{sec:Theory},
we study the decay of a neutral pion ($\pi^0 $) of \mbox{mass $M>0$}  
into two nonstandard photons ($\widetilde{\gamma}$), with
energies and 3-momenta denoted as follows:
\beq
\pi^0 (E,\,\vec{q}\,)\to \widetilde{\gamma}(\omega,\,\vec{k}\,)+\widetilde{\gamma}(\omega',\,\vec{k}'\,)\,,
\eeq
for on-shell energies $E(\vec{q}\,)=\sqrt{|\vec{q}\,|^2+M^2}$
and $\omega(\vec{k}\,)\geq0$ from \eqref{eq:disp-rel-effective-mass-square}.
This Lorentz-violating decay process has already been discussed qualitatively
in Ref.~\citen{ColemanGlashow1999}. A later heuristic discussion
in terms of effective-mass-squares~\cite{KaufholdKlinkhamer2005}
has been given in App.~A of Ref.~\citen{DKR2016}.

Lorentz-violating decays have been discussed extensively
in Ref.~\citen{KaufholdKlinkhamer2005}.
The general expression for the neutral-pion decay
parameter $\gamma$ is then as follows
(see, e.g., Sec.~3.6 of Ref.~\citen{DeWitSmith1986}):
\beqa\label{eq:gamma}
\gamma(\vec{q}\,)
&\equiv&
2\,E(\vec{q}\,)\;\Gamma(\vec{q}\,) 
\nonumber\\[1mm]
&=&
\frac{1}{2}\,\frac{1}{(2\pi)^2}\,
\int \frac{d^3k}{2\,\omega(\vec{k}\,)}\,
\int \frac{d^3k'}{2\,\omega(\vec{k}')}\,
\delta^4(q-k-k')\;|\mathcal{M}|^2\,,
\eeqa
with the symmetry factor $1/2$ and
the matrix element $\mathcal{M}$.
In a Lorentz-invariant theory, $|\mathcal{M}|^2$ is a scalar
and the right-hand side of \eqref{eq:gamma}
is manifestly Lorentz invariant, so that
$\gamma$ becomes independent of $\vec{q}$ and is called
the decay constant.

Following the discussion of Secs.~4.5 and 7.1 in Ref.~\citen{DeWitSmith1986},
we now take the effective pion-photon-photon interaction to be
given by
\beq
\mathcal{L}_\text{eff}=\alpha\,C\,\phi\,
\epsilon^{\alpha\beta\gamma\delta}\,F_{\alpha\beta}\,F_{\gamma\delta}\,,
\eeq
with the neutral-pion pseudoscalar field $\phi$, the Maxwell field strength
$F_{\mu\nu}\equiv\partial_\mu A_\nu-\partial_\nu A_\mu$, and a
coupling constant $C$ of mass dimension $-1$.
A straightforward but tedious calculation of the Lorentz-violating decay parameter \eqref{eq:gamma} then gives
\beqa\label{eq:gamma-result}
\gamma(\vec{q}\,)
&=&
\frac{1}{2}\,\frac{1}{8\pi}\,\;\big(8\,\alpha\, C\big)^2\,\frac{1}{2}\,
M^4 \; g\left(\frac{q}{M},\,\kappa\right) \,,
\eeqa
in terms of a dimensionless function $g$ of the
3-momentum modulus $q\equiv |\vec{q}\,|$ in units of $M$
and the Lorentz-violating parameter $\kappa$.
The calculated function $g(q/M,\,\kappa)$ has the following
properties:
\bsubeqs\label{eq:g-properties}
\beqa
g(q/M,\,0) &=&  1\,,
 \\[2mm]
g(0,\,\kappa) &=&
\frac{{\sqrt{1 - {\kappa}^2}}}{{\left( 1 - \kappa \right) }^3}=
1+3\,\kappa + \text{O}(\kappa^2) \,,
 \\[2mm]
g(q/M,\,\kappa)  &=&  0\,, 
\;\;\;\;\text{for}\;\;
\sqrt{q^2+M^2} \geq E^{\,(\text{cutoff})}\,,
\eeqa
\esubeqs
with cutoff energy
\beq\label{eq:E-cutoff}
E^{\,(\text{cutoff})}
=    \sqrt{\frac{1-\kappa}{-2\,\kappa}}\;M
\sim \frac{M}{\sqrt{-2\,\kappa}}\;.
\eeq
The last approximate expression for the cutoff energy in \eqref{eq:E-cutoff}
agrees with the kinematic result of Ref.~\citen{ColemanGlashow1999}.

\begin{figure}[t]   
\centering
\includegraphics[width=0.6\textwidth]{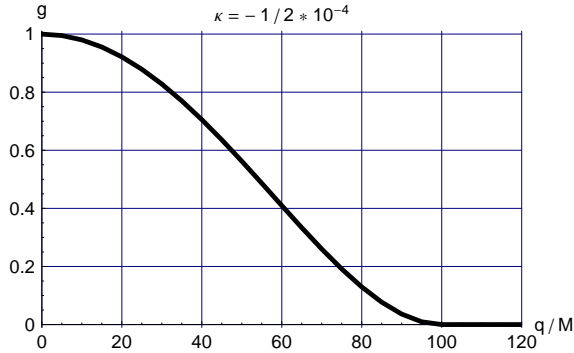}
\caption{Function $g(q/M,\,\kappa)$ from \eqref{eq:g-simpleqst}
for the neutral-pion decay parameter \eqref{eq:gamma-result}.
The numerical value of the Lorentz-violating parameter $\kappa$ is taken to be $\kappa=-5 \times 10^{-5}$, which gives a
cutoff momentum $q_{c} \sim 10^{2}\,M$, according to \eqref{eq:g-c}.
The shape of the function $g$ shown is close to the asymptotic
shape $(1-x^2)^2$, with definition $x \equiv q/q_{c}$.
} \label{Fig:g-function}
\vspace*{0cm}
\end{figure}

The exact formula for $g(q/M,\,\kappa)$ is, at first, rather cumbersome
but can be brought to a manageable form,
\bsubeqs\label{eq:g-simpleqst}
\beqa
g\left(q/M,\,\kappa\right) &=&
\left\{
\begin{array}{ll}
\displaystyle{\frac{{\sqrt{1-{\kappa}^2}}}{{\left(1-\kappa \right) }^3}}\;
\Big[1-(q/q_{c})^2\,\Big]^2
& \;\text{for}\;\;  q < q_{c}\,,\\[4mm]
0
& \;\text{for}\;\;  q \geq q_{c}\,,
\end{array}
\right.
\\[2mm]
\label{eq:g-c}
q_{c} &=& \sqrt{\frac{1+\kappa}{-2\,\kappa}}\;M
\sim \frac{M}{\sqrt{-2\,\kappa}}\;.
\eeqa
\esubeqs
Figure~\ref{Fig:g-function} gives a plot of the
function $g(q/M,\,\kappa)$ for a relatively large absolute  value of
the negative Lorentz-violating parameter $\kappa$.

The final result \eqref{eq:g-simpleqst}
for the Lorentz-violating neutral-pion decay parameter \eqref{eq:gamma-result}
can be used in numerical simulations of extensive air showers,
as discussed in Ref.~\citen{DKR2016}.

\section{Discussion}
\label{sec:Discussion}
\noindent
The first version of this paper dates from October, 2016.
Since then, numerical simulations of extensive air showers
have been performed~\cite{KNR2017}, which include the effects
of two Lorentz-violating decay processes in the
theory considered, photon decay
into an electron-positron pair as calculated in Ref.~\citen{KS2008}
and modified neutral-pion decay into two photons
as calculated in the present paper.
Comparing the simulated values of
the average atmospheric depth of the shower maximum
$\left<X_\text{max}\right>$ to the measured values from
the Pierre Auger Observatory gives a new bound on the negative
Lorentz-violating parameter $\kappa$, namely,
$\kappa > -3 \times 10^{-19} \, (98\%\, \text{CL})$.

Remark that the numerical value of this new negative $\kappa$ bound
is of the same order as the qualitative bound
$0 \leq v_\gamma-v_{\pi^{0}} < 10^{-20}$
from Ref.~\citen{Antonov-etal2001}.
The bound of Ref.~\citen{Antonov-etal2001} relies,
however, on a sharp kinematic cutoff~\cite{ColemanGlashow1999}
of the standard neutral-pion decay rate and does not take possible
photon-decay effects into account.
The Lorentz violation considered in Ref.~\citen{Antonov-etal2001}
does not trace back to a
consistent theory of elementary particle interactions,
whereas the bound of Ref.~\citen{KNR2017} follows from
explicit decay rates calculated in standard
quantum electrodynamics (QED)
with a single Lorentz-violating term added to the photonic action.

Recall, finally, that the isotropic Lorentz-violating term
in the photon sector can be moved into the fermion sector
by an appropriate coordinate transformation;
see App.~B of Ref.~\citen{KS2008} for details and
further references.
In fact, the Lorentz-violating parameter $\kappa$ measures the
relative difference in the photon phase velocity
and the maximum attainable velocity
of the massive Dirac fermions considered
(quarks and leptons), as clarified by
Eq.~(4) in Ref.~\citen{KNR2017}.

\vspace*{-3mm}
\section*{Acknowledgments}
\vspace*{-0mm}
\noindent
It is a pleasure to thank M. Risse for many discussions over the last years
and the referee for useful comments.
This work was supported in part by the German Research Foundation (DFG)
under \mbox{Grant No. KL 1103/4-1.}



\end{document}